\documentclass[a4paper,fleqn]{article}
\usepackage{amsmath}

\begin{document}

\title{\textbf{On integrability of a third-order complex nonlinear wave equation}}

\author{\textsc{Sergei Sakovich}\bigskip \\
\small Institute of Physics, National Academy of Sciences of Belarus \\
\small sergsako@gmail.com}

\date{}

\maketitle

\begin{abstract}
We show that the new third-order complex nonlinear wave equation, introduced recently by M\"{u}ller-Hoissen [arXiv:2202.04512], does not pass the Painlev\'{e} test for integrability. We find two reductions of this equation, one integrable and one non-integrable, whose solutions jointly cover all solutions of the original equation.
\end{abstract}

\section{Introduction}

The following new third-order complex nonlinear wave equation was introduced recently by M\"{u}ller-Hoissen \cite{MH22}:
\begin{equation}
\left( \frac{f_{xt}}{f} \right)_t + 2 \left( f f^* \right)_x = 0 , \label{e1}
\end{equation}
where $f(x,t)$ is a complex function of two real variables, subscripts denote respective derivatives, and the asterisk stands for the complex conjugate. This nonlinear wave equation \eqref{e1} was called a completely integrable partial differential equation in \cite{MH22}, and its multi-soliton solutions were obtained there.

Let us note, however, that this new nonlinear equation was studied in \cite{MH22} not in its original form \eqref{e1} but in the form of the following system of two equations:
\begin{equation}
a_t = \left( f f^* \right)_x , \qquad f_{xt} + 2 a f = 0 , \label{e2}
\end{equation}
where $a(x,t)$ is a real function. Being the first ``negative flow'' of the nonlinear Schr\"{o}dinger equation's hierarchy, this nonlinear system \eqref{e2} is integrable and possesses multi-soliton solutions. The integrable system \eqref{e2} is obviously a reduction of the new nonlinear wave equation \eqref{e1}, in the sense that all solutions of \eqref{e2} (except for $f=0$, of course) are solutions of \eqref{e1} as well, because \eqref{e1} follows from \eqref{e2} through elimination of the dependent variable $a$. In other words, the new equation \eqref{e1} possesses multi-soliton solutions because it possesses an integrable reduction. But is the new equation \eqref{e1} integrable itself? No Lax pair of \eqref{e1} is known.

In the present paper, we study the integrability of the new equation \eqref{e1} by means of the singularity analysis (a.k.a. the Painlev\'{e} analysis) in its version for partial differential equations \cite{WTC83,T89}. In Section~\ref{s2}, we show that the nonlinear equation \eqref{e1} does not pass the Painlev\'{e} test for integrability. In Section~\ref{s3}, we find two reductions of the nonlinear equation \eqref{e1}, one integrable and one non-integrable, whose solutions jointly cover all solutions of the original equation. Section~\ref{s4} contains concluding remarks.

\section{Singularity analysis} \label{s2}

In our experience, the Painlev\'{e} analysis is a reliable and convenient method to test the integrability of nonlinear wave equations, including high-order, non-evolutionary, multi-component and high-dimensional ones \cite{S94a,S94b,S95,S97,S98,KS01,KSY03,KS05,S05,S08,S11,S13,S17,S18,S19,S21}. The reliability of the Painlev\'{e} test for integrability has been empirically verified by the analysis of wide classes of nonlinear equations, such as fifth-order Korteweg--de~Vries type equations \cite{HO85}, bilinear equations \cite{GRH90}, coupled Korteweg--de~Vries equations \cite{K97,S99,S01,S14}, coupled higher-order nonlinear Schr\"{o}dinger equations \cite{ST00}, generalized Ito equations \cite{KKS01}, sixth-order nonlinear wave equations \cite{KKSST08}, seventh-order Korteweg--de~Vries type equations \cite{X14}, etc.

To start the singularity analysis of the nonlinear equation \eqref{e1}, we rewrite this equation and its complex conjugate
\begin{equation}
\left( \frac{f_{xt}^*}{f^*} \right)_t + 2 \left( f f^* \right)_x = 0 \label{e3}
\end{equation}
as the following system of two polynomial equations:
\begin{gather}
f f_{xtt} - f_t f_{xt} + 2 f^2 ( f g )_x = 0 , \notag \\
g g_{xtt} - g_t g_{xt} + 2 g^2 ( f g )_x = 0 , \label{e4}
\end{gather}
where $g$ stands for $f^*$. From now on, we consider $f(x,t)$ and $g(x,t)$ as two mutually independent complex functions of two complex variables $x$ and $t$.

A singularity manifold $\phi (x,t) = 0$ is non-characteristic for this system \eqref{e4} if $\phi_x \phi_t \ne 0$, and we take $\phi_t = 1$ without loss of generality,
\begin{equation}
\phi = t + \psi (x) , \qquad \psi_x \ne 0 , \label{e5}
\end{equation}
where $\psi (x)$ is arbitrary. Substitution of the expansions
\begin{gather}
f = f_0 (x) \phi^p + \dotsb + f_r (x) \phi^{p+r} + \dotsb , \notag \\
g = g_0 (x) \phi^q + \dotsb + g_r (x) \phi^{q+r} + \dotsb \label{e6}
\end{gather}
to the nonlinear system \eqref{e4} determines the leading exponents $p$ and $q$ (i.e., the dominant behavior of solutions $f$ and $g$ near the singularity manifold $\phi = 0$) and the resonances $r$ (i.e., the positions, where arbitrary functions can enter the expansions). In this way, we find the following one branch to study further:
\begin{equation}
p = q = -1 , \qquad r = -1, 0, 2, 2, 3, 4 , \label{e7}
\end{equation}
where $r = -1$ corresponds to the arbitrariness of $\psi (x)$ in \eqref{e5}.

For this branch \eqref{e7}, we try to represent the general solution of \eqref{e4} by the Laurent type expansions
\begin{equation}
f = \sum_{i=0}^{\infty} f_i (x) \phi^{i-1} , \qquad g = \sum_{i=0}^{\infty} g_i (x) \phi^{i-1} , \label{e8}
\end{equation}
with $\phi$ given by \eqref{e5}. We substitute \eqref{e8} to \eqref{e4}, collect terms with $\phi^{n-5}$, for $n=0,1,2,3, \dotsc$ separately, and obtain in this way the following.

For $n=0$, where we have a single resonance due to \eqref{e7}, we get the expression for $g_0$,
\begin{equation}
g_0 = - \frac{1}{f_0} , \label{e9}
\end{equation}
while the function $f_0 (x)$ remains arbitrary.

For $n=1$, there is no resonance, and we get the expressions
\begin{equation}
f_1 = - \frac{{f_0}'}{2 \psi'} , \qquad g_1 = - \frac{{f_0}'}{2 f_0^2 \psi'} , \label{e10}
\end{equation}
where the prime denotes the derivative with respect to $x$.

For $n=2$, we have a double resonance, the functions $f_2 (x)$ and $g_2 (x)$ remain arbitrary, the compatibility conditions are identically satisfied by \eqref{e9} and \eqref{e10}, and no restrictions for $\psi (x)$ and $f_0 (x)$ appear.

For $n=3$, where we have a single resonance, we get the expression for $g_3$,
\begin{gather}
g_3 = \frac{f_3}{f_0^2} - \frac{g_2 {f_0}'}{2 f_0 \psi'} + \frac{\left( {f_0}' \right)^3}{2 f_0^4 (\psi')^3} + \frac{{f_2}'}{2 f_0^2 \psi'} \notag \\
\qquad - \frac{{g_2}'}{\psi'} + \frac{\left( {f_0}' \right)^2 \psi''}{2 f_0^3 (\psi')^4} - \frac{{f_0}' {f_0}''}{2 f_0^3 (\psi')^3} , \label{e11}
\end{gather}
the function $f_3 (x)$ remains arbitrary, but the following nontrivial compatibility condition appears:
\begin{equation}
\left( \frac{f_2}{f_0}  + f_0 g_2 \right)' = 0 . \label{e12}
\end{equation}

The fact that we have got a nontrivial compatibility condition at a resonance means that we have to modify our expansions of solutions by additional logarithmic terms. Consequently, the nonlinear equation \eqref{e1} does not pass the Painlev\'{e} test for integrability.

\section{Two reductions} \label{s3}

Let us return to the nonlinear equation \eqref{e1} and consider it together with its complex conjugate \eqref{e3}. From these two equations, we get the relation
\begin{equation}
\left( \frac{f_{xt}}{f} - \frac{f_{xt}^*}{f^*} \right)_t = 0 \label{e13}
\end{equation}
satisfied by any solution of \eqref{e1}. We introduce two real functions of two real variables, $u(x,t)$ and $v(x,t)$, such that
\begin{equation}
\frac{f_{xt}}{f} = u(x,t) + \mathrm{i} v(x,t) , \label{e14}
\end{equation}
where $\mathrm{i}^2 = -1$. Then, it follows from \eqref{e13} and \eqref{e14} that $v_t = 0$, and \eqref{e14} takes the form
\begin{equation}
\frac{f_{xt}}{f} = u(x,t) + \mathrm{i} w(x) , \label{e15}
\end{equation}
where $w(x)$ is a real function of one real variable.

It was pointed out in \cite{MH22} that the nonlinear equation \eqref{e1} is invariant under the transformation
\begin{equation}
x \mapsto h(x) , \label{e16}
\end{equation}
where $h(x)$ is an arbitrary function. In other words, if a function $f(x,t)$ is a solution of the nonlinear equation \eqref{e1}, then $f(h(x),t)$ is also a solution of \eqref{e1}, for any function $h(x)$.

Under the transformation \eqref{e16}, the function $w(x)$ in the relation \eqref{e15} changes in the following way:
\begin{equation}
w(x) \mapsto h'(x) \, w(h(x)) , \label{e17}
\end{equation}
where the prime denotes the derivative. Therefore, for any solution $f(x,t)$ which corresponds to any nonzero function $w(x)$ in \eqref{e15}, the integral
\begin{equation}
x = \int w(h) \, dh \label{e18}
\end{equation}
determines (at least, locally) the function $h(x)$ of the transformation \eqref{e16}, such that the transformed solution $f(h(x),t)$ corresponds to $w=1$. Consequently, we can use the relation \eqref{e15} in the form
\begin{equation}
\frac{f_{xt}}{f} = u(x,t) + \mathrm{i} k , \qquad k = 0, 1 , \label{e19}
\end{equation}
provided that, in the case of $k=1$, every single solution $f(x,t)$ represents the whole its equivalence class $f(h(x),t)$ with any function $h(x)$.

The original nonlinear equation \eqref{e1} together with the relation \eqref{e19} give us the following system of two equations:
\begin{gather}
f_{xt} - ( u + \mathrm{i} k ) f = 0 , \notag \\
u_t + 2 \left( f f^* \right)_x = 0 , \qquad k = 0, 1 , \label{e20}
\end{gather}
where $f(x,t)$ is a complex function of two real variables, but $u(x,t)$ is a real function of two real variables. The two cases of the system \eqref{e20}, with $k=0$ and with $k=1$, are two distinct reductions of the original equation \eqref{e1}. Solutions of these two reductions jointly cover all solutions of \eqref{e1}, provided that every solution $f(x,t)$ of the system \eqref{e20} with $k=1$ is generalized as $f(h(x),t)$, with any function $h(x)$. (Note that the system \eqref{e20} with $k=1$ is not invariant under the transformation \eqref{e16}, whereas the system \eqref{e20} with $k=0$ is invariant.) The system \eqref{e20} with $k=0$ is just the system \eqref{e2} studied in \cite{MH22}, the correspondence being $u = -2 a$. The system \eqref{e20} with $k=1$ is something new.

Let us apply the Painlev\'{e} test for integrability to the system \eqref{e20}. We combine \eqref{e20} with its complex conjugate, denote $f^*$ as $g$, treat $f$ and $g$ as mutually independent, complexify all variables, and consider the following system of three equations:
\begin{gather}
f_{xt} - ( u + \mathrm{i} k ) f = 0 , \notag \\
g_{xt} - ( u - \mathrm{i} k ) g = 0 , \notag \\
u_t + 2 \left( f g \right)_x = 0 , \qquad k = 0, 1 , \label{e21}
\end{gather}
where $f(x,t)$, $g(x,t)$ and $u(x,t)$ are complex function of two complex variables.

We find from \eqref{e21} that, near any non-characteristic singularity manifold $\phi (x,t) = 0$ with $\phi$ given by \eqref{e5} and arbitrary $\psi (x)$, the leading exponents of $f$, $g$ and $u$ are $-1$, $-1$ and $-2$, respectively, and the resonances are
\begin{equation}
r = -1, 0, 2, 3, 4 , \label{e22}
\end{equation}
where $r = -1$ corresponds to the arbitrariness of $\psi (x)$ in \eqref{e5}. Then we substitute the expansions
\begin{gather}
f = \sum_{i=0}^{\infty} f_i (x) \phi^{i-1} , \qquad g = \sum_{i=0}^{\infty} g_i (x) \phi^{i-1} , \notag \\
u = \sum_{i=0}^{\infty} u_i (x) \phi^{i-2} \label{e23}
\end{gather}
to the system \eqref{e21}, collect terms with $\phi^{n-3}$, for $n=0,1,2,3, \dotsc$ separately, and obtain in this way the following.

For $n=0$, we get the expressions
\begin{equation}
g_0 = - \frac{1}{f_0} , \qquad u_0 = 2 \psi' , \label{e24}
\end{equation}
the function $f_0 (x)$ remains arbitrary, and no nontrivial compatibility condition appears at this resonance.

For $n=1$, where we have no resonance, we get the expressions
\begin{equation}
f_1 = - \frac{{f_0}'}{2 \psi'} , \qquad g_1 = - \frac{{f_0}'}{2 f_0^2 \psi'} , \qquad u_1 = 0 . \label{e25}
\end{equation}

For $n=2$, we get the expressions
\begin{equation}
f_2 = - \frac{f_0 ( u_2 + \mathrm{i} k )}{2 \psi'} , \qquad g_2 = \frac{u_2 - \mathrm{i} k}{2 f_0 \psi'} , \label{e26}
\end{equation}
the function $u_2 (x)$ remains arbitrary, and no nontrivial compatibility condition appears at this resonance.

For $n=3$, we get the expressions
\begin{gather}
u_3 = - \frac{{u_2}'}{2 \psi'} + \frac{( u_2 + \mathrm{i} k ) \psi''}{2 (\psi')^2} , \notag \\
g_3 = \frac{f_3}{f_0^2} - \frac{\mathrm{i} k {f_0}'}{2 f_0^2 (\psi')^2} + \frac{\left( {f_0}' \right)^3}{2 f_0^4 (\psi')^3} - \frac{3 {u_2}'}{4 f_0 (\psi')^2} \notag \\
\qquad - \frac{\mathrm{i} k \psi''}{4 f_0 (\psi')^3} + \frac{3 u_2 \psi''}{4 f_0 (\psi')^3} + \frac{\left( {f_0}' \right)^2 \psi''}{2 f_0^3 (\psi')^4} - \frac{{f_0}' {f_0}''}{2 f_0^3 (\psi')^3} , \label{e27}
\end{gather}
and the function $f_3 (x)$ remains arbitrary. However, the following compatibility condition appears at this resonance:
\begin{equation}
k \psi'' = 0 , \label{e28}
\end{equation}
which is not satisfied identically if $k=1$. Consequently, the system \eqref{e20} with $k=1$ does not pass the Painlev\'{e} test for integrability.

If $k=0$, the compatibility condition \eqref{e28} is satisfied identically, and we continue computations. For $n=4$, we obtain explicit expressions for $u_4$ and $g_4$ (cumbersome ones, therefore omitted here), the function $f_4 (x)$ remains arbitrary, and no nontrivial compatibility condition appears at this resonance. Consequently, the system \eqref{e20} with $k=0$ has passed the Painlev\'{e} test for integrability (and this is an expected result, of course).

\section{Conclusion} \label{s4}

In this paper, we studied the integrability of the third-order complex nonlinear wave equation \eqref{e1}, introduced recently in \cite{MH22}. We used the Painlev\'{e} test for integrability and obtained the following results.

The nonlinear equation \eqref{e1} fails to pass the Painlev\'{e} test for integrability. Therefore we believe that one cannot find any good Lax pair for this equation.

The nonlinear equation \eqref{e1} possesses two reductions, one integrable and one non-integrable, whose solutions jointly cover all solutions of this equation.

The integrable reduction is the system \eqref{e20} with $k=0$, studied in \cite{MH22} in the form \eqref{e2}. This reduction corresponds to the condition $f_{xt} / f = f_{xt}^* / f^*$ imposed on solutions of \eqref{e1}. This integrable reduction is just what provides the original equation \eqref{e1} with multi-soliton solutions.

The non-integrable reduction is the system \eqref{e20} with $k=1$. It corresponds to the condition $f_{xt} / f \ne f_{xt}^* / f^*$ imposed on solutions of \eqref{e1}. Solutions of this non-integrable reduction must be generalized by the arbitrary coordinate transformation \eqref{e16}, in order to cover all those solutions of \eqref{e1} which are not solutions of the integrable reduction.

It seems to be an interesting future problem to find any explicit solutions of the new (non-integrable) system \eqref{e20} with $k=1$. We believe that there are no multi-soliton solutions, and that the travelling waves (if there are any) interact inelastically.


\begin{thebibliography}{99}

\small

\bibitem{MH22} F. M\"{u}ller-Hoissen, A vectorial binary Darboux transformation for the first member of the negative part of the AKNS hierarchy, arXiv:2202.04512v3.

\bibitem{WTC83} J. Weiss, M. Tabor, G. Carnevale, The Painlev\'{e} property  for partial differential equations, J. Math. Phys. 24 (1983) 522--526.

\bibitem{T89} M. Tabor, Chaos and Integrability in Nonlinear Dynamics: An Introduction, Wiley, New York, 1989.

\bibitem{S94a} S.Yu. Sakovich, Painlev\'{e} analysis and B\"{a}cklund transformations of Doktorov--Vlasov equations, J. Phys. A: Math. Gen. 27 (1994) L33--L38.

\bibitem{S94b} S.Yu. Sakovich, Painlev\'{e} analysis of new soliton equations by Hu, J. Phys. A: Math. Gen. 27 (1994) L503--L505.

\bibitem{S95} S.Yu. Sakovich, On zero-curvature representations of evolution equations, J. Phys. A: Math. Gen. 28 (1995) 2861--2869.

\bibitem{S97} S.Yu. Sakovich, Painlev\'{e} analysis of a higher-order nonlinear Schr\"{o}dinger equation, J. Phys. Soc. Jpn. 66 (1997) 2527--2529.

\bibitem{S98} S.Yu. Sakovich, On integrability of a $(2+1)$-dimensional perturbed KdV equation, J. Nonlinear Math. Phys. 5 (1998) 230--233; arXiv:solv-int/9805012.

\bibitem{KS01} A. Karasu-Kalkanl\i, S.Yu. Sakovich, B\"{a}cklund transformation and special solutions for the Drinfeld--Sokolov--Satsuma--Hirota system of coupled equations, J. Phys. A: Math. Gen. 34 (2001) 7355--7358; arXiv:nlin/0102001.

\bibitem{KSY03} A. Karasu-Kalkanl\i, S.Yu. Sakovich, \'{I}. Yurdu\c{s}en, Integrability of Kersten--Krasil'shchik coupled KdV--mKdV equations: singularity analysis and Lax pair, J. Math. Phys. 44 (2003) 1703--1708; arXiv:nlin/0206046.

\bibitem{KS05} A. Karasu-Kalkanl\i, S. Sakovich, Singularity analysis of a spherical Kadomtsev--Petviashvili equation, J. Phys. Soc. Jpn. 74 (2005) 505--507; arXiv:nlin/0404037.

\bibitem{S05} S. Sakovich, Enlarged spectral problems and nonintegrability, Phys. Lett. A 345 (2005) 63--68; arXiv:nlin/0504037.

\bibitem{S08} S. Sakovich, Integrability of the vector short pulse equation, J. Phys. Soc. Jpn. 77 (2008) 123001; arXiv:0801.3179.

\bibitem{S11} S. Sakovich, Singularity analysis and integrability of a Burgers-type system of Foursov, Symmetry Integr. Geom. Methods Appl. 7 (2011) 002; arXiv:1010.5709.

\bibitem{S13} S. Sakovich, On two aspects of the Painlev\'{e} analysis, Int. J. Analysis 2013 (2013) 172813; arXiv:solv-int/9909027.

\bibitem{S17} S. Sakovich, Integrability study of a four-dimensional eighth-order nonlinear wave equation, Nonlinear Phenom. Complex Syst. 20 (2017) 267--271; arXiv:1607.08408.

\bibitem{S18} S. Sakovich, On a new avatar of the sine-Gordon equation, Nonlinear Phenom. Complex Syst. 21 (2018) 62--68; arXiv:1703.04678.

\bibitem{S19} S. Sakovich, A new Painlev\'{e}-integrable equation possessing KdV-type solitons, Nonlinear Phenom. Complex Syst. 22 (2019) 299--304; arXiv:1907.01324.

\bibitem{S21} S. Sakovich, Integrability of one bilinear equation: singularity analysis and dimension, Nonlinear Phenom. Complex Syst. 24 (2021) 311--316; arXiv:2109.02073.

\bibitem{HO85} H. Harada, S. Oishi, A new approach to completely integrable partial differential equations by means of the singularity analysis, J. Phys. Soc. Jpn. 54 (1985) 51--56.

\bibitem{GRH90} B. Grammaticos, A. Ramani, J. Hietarinta, A search for integrable bilinear equations: The Painlev{\'e} approach, J. Math. Phys. 31 (1990) 2572--2578.

\bibitem{K97} A. Karasu-Kalkanl\i, Painlev\'{e} classification of coupled Korteweg--de~Vries systems, J. Math. Phys. 38 (1997) 3616--3622.

\bibitem{S99} S.Yu. Sakovich, Coupled KdV equations of Hirota--Satsuma type, J. Nonlinear Math. Phys. 6 (1999) 255--262; arXiv:solv-int/9901005.

\bibitem{S01} S.Yu. Sakovich, Addendum to: Coupled KdV equations of Hirota--Satsuma type, J. Nonlinear Math. Phys. 8 (2001) 311--312; arXiv:nlin/0104072.

\bibitem{S14} S. Sakovich, A note on the Painlev\'{e} property of coupled KdV equations, Int. J. Part. Diff. Eqns. 2014 (2014) 125821; arXiv:nlin/0402004.

\bibitem{ST00} S.Yu. Sakovich, T. Tsuchida, Symmetrically coupled higher-order nonlinear Schr\"{o}dinger equations: singularity analysis and integrability, J. Phys. A: Math. Gen. 33 (2000) 7217--7226; arXiv:nlin/0006004.

\bibitem{KKS01} A. Karasu-Kalkanl\i, A. Karasu, S.Yu. Sakovich, Integrability of a generalized Ito system: the Painlev\'{e} test, J. Phys. Soc. Jpn. 70 (2001) 1165--1166; arXiv:nlin/0102030.

\bibitem{KKSST08} A. Karasu-Kalkanl\i, A. Karasu, A. Sakovich, S. Sakovich, R. Turhan, A new integrable generalization of the Korteweg--de~Vries equation, J. Math. Phys. 49 (2008) 073516; arXiv:0708.3247.

\bibitem{X14} G.Q. Xu, The integrability for a generalized seventh-order KdV equation: Painlev\'{e} property, soliton solutions, Lax pairs and conservation laws, Phys. Scr. 89 (2014) 125201.

\end{thebibliography}
\end{document}